\documentclass[prb,preprint]{revtex4}

\pdfoutput=1

\usepackage{graphicx}

\begin{document}

\title{Fermi arc, pseudogap and strange-metal phase 
\linebreak in hole-dopd lanthanum cuprates}

\medskip 

\date{April 16, 2017} \bigskip

\author{Manfred Bucher \\}
\affiliation{\text{\textnormal{Physics Department, California State University,}} \textnormal{Fresno,}
\textnormal{Fresno, California 93740-8031} \\}

\begin{abstract}
Hole doping of $La_{2-x}Ae_xCuO_4$ ($Ae=Sr,Ba$) and $La_{2-y-x}Ln_ySr_{x}CuO_4$ ($Ln = Nd, Eu; \, y = 0.4, \, 0.2$) introduces unidirectional charge density waves (CDWs) of incommensurability $\delta_c(x)$ in domains of the $CuO_2$ planes. A periodic structure, each CDW gives rise to a Bragg-reflection mirror of extension $\delta_c(x)$ that attaches to a nodal point \.{Q} on the planar diagonal in reciprocal space. This confines itinerant holes to a Fermi arc about  \.{Q}, leaving a pseudogap along the remainder of the underlying Fermi surface. The length of the Fermi arc and the magnitude of the pseudogap both are determined by $\delta_c(x)$. The pseudogap closes when the Fermi arc reaches the antinodal symmetry points M.
This is the case at a doping level $x^*_0 = 0.182$ for $La_{2-x}Ae_{x}CuO_{4}$ at $T=0$ (quantum critical point, QCP) and otherwise at a doping-dependent pseudogap temperature $T^*(x)$ that marks the boundary between the compounds' pseudogap phase and strange-metal phase. 
The different value of the observed QCP in $La_{2-y-x}Ln_ySr_{x}CuO_4$, $x^*_{0} = 0.235$, is attributed to extra magnetic order from $Ln^{3+}$ ions with a finite magnetic moment instead of $La^{3+}$ with none.
The possibility of quantum oscillations in $La_{2-y-x}Ln_ySr_{x}CuO_4$ in the high-end doping interval of their pseudogap phase, $0.182 < x < 0.235$, is raised. The strange-metal phase is interpreted as a consequence of conflicting Bragg reflection conditions for the crystals' itinerant charge carriers when boundaries of the BZ and the CDW mirrors coincide, frustrating \emph{umklapp} processes of carrier-carrier scattering.


\end{abstract}

\maketitle

\pagebreak

\section{CHARGE DENSITY WAVES OF DOPED HOLES}

Charge density waves (CDWs) in high-transition-temperature superconductors were first discovered indirectly by neutron diffraction in $La_{1.6-x}Nd_{0.4}Sr_{x}CuO_{4}$ and later confirmed with X-ray diffraction.\cite{1,2,3,4} They were subsequently observed indirectly by hard X-ray diffraction or directly by resonant soft X-ray scattering
in the isostructural (`214') compounds $La_{2-x}Ae_{x}CuO_{4}$ ($Ae = Sr, Ba$) and $La_{1.8-x}Eu_{0.2}Sr_{x}CuO_{4}$.\cite{5,6,7,8,9,10,11,12,13,14,15,16,17,18,19,20,21} 
In those materials the CDWs are frequently accompanied by magnetic dipole waves (MDWs), also called spin density waves (SDWs), observed by neutron scattering.\cite{22,23,24,25,26,27,28,29,30,31,32,33}
The combined occurrence of both density waves, called stripes,\cite{34}  characterizes a region in the $x$-$T$ phase diagram, shown in Fig. 1a  for $La_{2-x}Sr_{x}CuO_{4}$.
The stripes reside in the crystals' $CuO_2$ planes of lattice constants $a_0$, $b_0$ and are incommensurate to the lattice periodicity of the $Cu^{2+}$ and $O^{2-}$ ions.
The doping dependence of the incommensurability (being a wave number)  of the charge ($c$) or magnetic ($m$) density waves is, in reciprocal lattice units (r.l.u.),
\begin{equation}
\delta_{c,m}(x)  = w_{c,m} \frac{\Omega^{\pm}}{{4}}\sqrt {x - {x_{0}^N}}  \; ,
\end{equation}
\noindent with a wave-kind factor $w_c=2$ or $w_m=1$ and a stripe-orientation factor $\Omega^{+}=\sqrt{2}$ for $x > x_6 \equiv 2/6^2  \simeq 0.056$ when stripes are parallel to the $a$ or $b$ axis, but $\Omega^{-} = 1$ for $x < x_6$ when stripes are diagonal (see Fig. 1b). Here $x_0^N$ is the doping concentration where the N\'{e}el temperature vanishes, $T_N(x_0^N) = 0$. The density waves emerge at the doping level $x_{0}^N = x_{10}\equiv 2/10^2 = 0.02$ where three-dimensional antiferromagnetism (3D-AFM) collapses. The derivation of Eq. (1) is based on a partition of the $CuO_2$ plane by pairs of doped holes, incorporating the observed stripe orientation, here in tetragonal approximation, $a_0 = b_0$.\cite{35}

As is well-known, the periodicity of atomic positions in a crystal gives rise to diffraction of waves---be it light waves of X-ray photons caused by scattering off the atoms' electrons, be it matter waves of neutrons caused by scattering off the atoms' magnetic moments, or matter waves of the crystals' valence electrons caused by scattering off the charged ion cores---all subject to the Bragg reflection condition. A periodic structure, each density wave in $La_{2-x}Ae_{x}CuO_{4}$ (LACO), $La_{1.6-x}Nd_{0.4}Sr_{x}CuO_{4}$ (Nd-LSCO) and $La_{1.8-x}Eu_{0.2}Sr_{x}CuO_{4}$ (Eu-LSCO) \emph{also} gives rise to a corresponding diffraction. 
For hard X-rays the CDW diffraction manifests itself indirectly by very weak satellite peaks in the diffraction pattern from lattice diffraction.\cite{34} 
Likewise, for neutrons the MDW diffraction gives rise to magnetic satellite peaks.\cite{34} 
Resonant scattering of soft X-ray photons, say near the Cu-$L_3$ absorption edge 

\pagebreak

\includegraphics[width=4.28in]{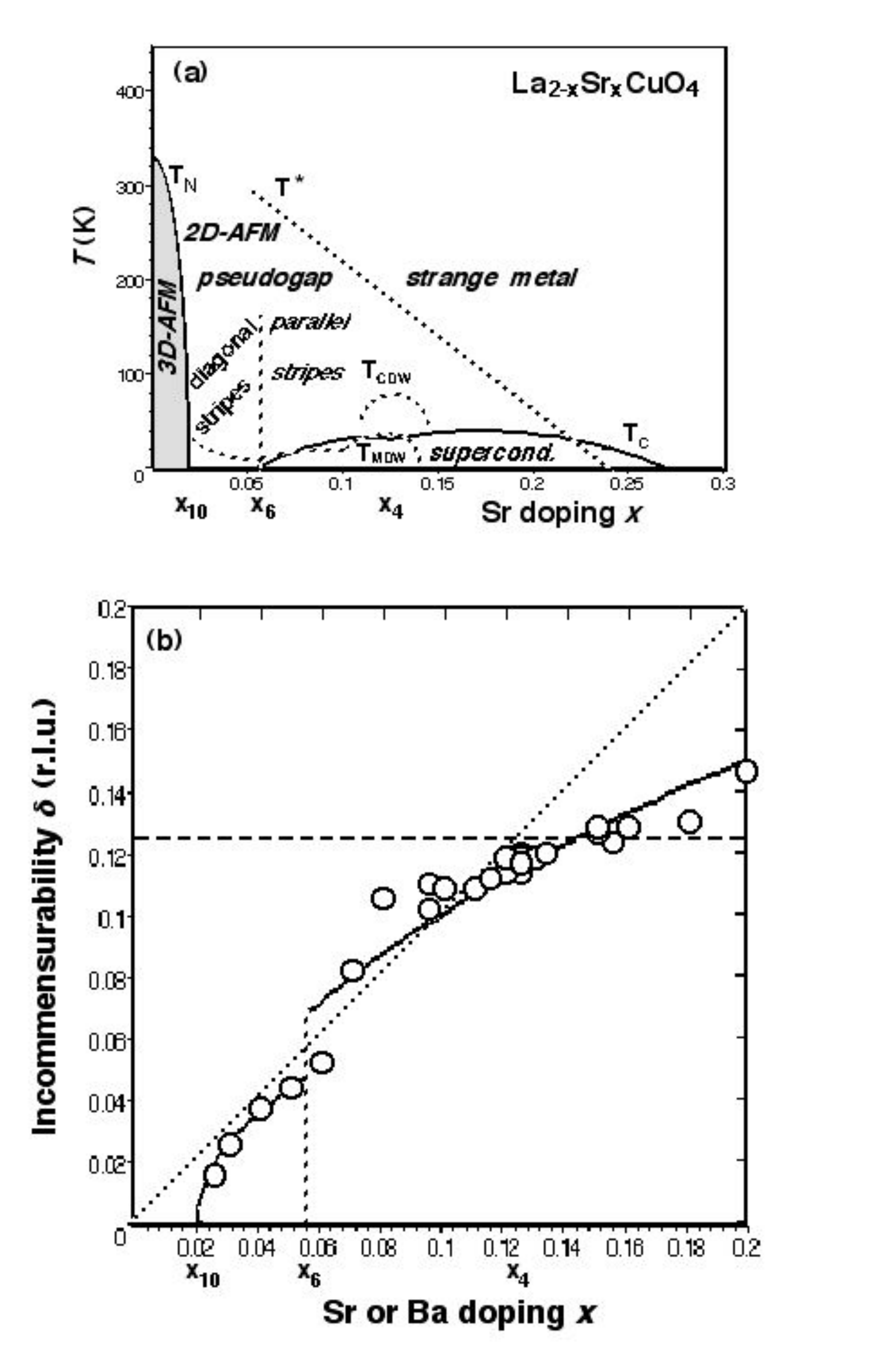}

\footnotesize 

\noindent FIG. 1. (a) Phase diagram of $La_{2-x}Sr_{x}CuO_{4}$ based on Refs. 22, 37-41, 55-60 with N\'{e}el temperature $T_N$, transition temperature of superconductivity $T_c$, pseudogap temperature $T^*$ (extrapolated to $T=0$) and temperatures $T_{CDW}$ and $T_{MDW}$ beneath which charge-density waves and, respectively, magnetic dipole waves are detected.
Special concentrations  $x_n \equiv 2/n^2$, marked on the ordinate axis, correspond to commensurate doping of one $Sr$ atom per $na_{0}\times nb_{0}$ area in each $LaO$ plane sandwiching a $CuO_2$ plane.  
\linebreak (b) Incommensurability of MDWs, $\delta_m^p = \delta$, and of CDWs $\delta_c^p = 2\delta$, in $La_{2-x}Ae_{x}CuO_{4}$ due to doping with $Ae = Sr$ or $Ba$, as well as in $La_{1.6-x}Nd_{0.4}Sr_{x}CuO_{4}$ and $La_{1.8-x}Eu_{0.2}Sr_{x}CuO_{4}$. Circles show data from neutron scattering or X-ray diffraction (Refs. 1-33). The broken solid curve is a graph of Eq. (1). Dotted slanted line $\delta = x$, dashed horizontal line at $\delta = 0.125 = 1/8$.
\pagebreak

\normalsize 

\noindent ($2p_{3/2} \rightarrow 3d$ transition, $h \nu \simeq
931.5$ eV), is found to be enhanced by charge modulations involving the valence electrons in the $CuO_2$ planes.\cite{36} This permits a direct probing of CDWs.\cite{6,7,8,9,17,21}

The sensitivity of detection of CDWs by hard X-ray diffraction or resonant soft X-ray scattering, as well as of MDWs by neutron scattering, differs in several respects. In the parallel regime, $x > x_6$, the temperature profiles beneath which CDWs and MDWs are detected are dome-like with maxima at $x_4 \equiv 2/4^2 = 1/8$, ranking $T_{CDW}(x) > T_c(x) > T_{MDW}(x) $ (see Fig. 1a). 
In the diagonal regime, $x < x_6$, only MDWs have been observed, no CDWs. 
The reasons for these differences are not entirely clear. Distinguishing between detection and existence of CDWs and MDWs, we want to \emph{posit} that combined CDWs/MDWs (stripes) always \emph{exist} in cuprates of the 
`214' family when hole-doped, regardless of their present ``visibility'' by X-rays or neutrons. Assuming their wave-number incommensurabiliy $\delta_{c,m}(x)$ given by Eq. (1), we want to consider the effect of diffraction by CDWs on the \emph{valence electrons} in the $CuO_2$ planes. 

\section{FERMI ARC}

Electrons in a crystal occupy quantum states according to their momentum, $\textbf{p}  = \hbar \textbf{k}$, with energy $\epsilon(\textbf{k})$. 
In a metal at $T = 0$, the occupied quantum states of the common highest ground-state energy $\hat{\epsilon}$ establish a Fermi surface, $\epsilon_F = \hat{\epsilon}$, in momentum space. Contrarily, in an insulator where $\hat{\epsilon} < \epsilon_F$, the Fermi surface remains unoccupied. Here the term ``surface'' is understood in a generalized sense: It is a genuine surface in 3D (spherical for free electrons), but a \emph{circumference} in 2D (circular for free charge carriers) if metallicity exists only in isolated planes of the crystal.
When the influence of the periodic crystal lattice is taken into account, the momentum space (or ``k-space,'' $\mathbf{k}  = \mathbf{p}/\hbar$) that houses the quantum states is divided into Brillouin zones (BZs) by the reciprocal lattice. Because of Bragg reflection of electron waves at BZ boundaries, quantum states in the second (or higher) extended BZ can be backfolded into the first BZ (called ``reduced BZ scheme''). In the ensuing treatment of the $CuO_{2}$ planes we consider only the in-plane components of their quantum state, $k_a$ and $k_b$. Also, since the following discussion involves incommensurabilities (reciprocal wavelengths), it is more convenient to address quantum states in reciprocal space, or ``$q$-space,'' $\mathbf{q} \equiv \mathbf{k}/2\pi$, rather than in $k$-space.

A stoichiometric $LaCuO_{4}$ crystal is a Mott insulator with 3D-AFM. In doped $La_{2-x}Ae_{x}CuO_{4}$, at a small doping level $x_{0}^N=0.02$, 3D-AFM collapses and a ``metallic'' (itinerant charge carrier) quantum state emerges at a certain position $\dot{Q} = (\dot{q},\dot{q})$ on the 2D $q$-space diagonal. It is surrounded by all other quantum states that are still insulating in the 2D-AFM phase on account of an energy gap.\cite{39} For easy reference we want to call $\dot{Q}$ the ``Fermi dot.'' Photoemission spectroscopy (ARPES) and Hubbard-model based calculations show that in LACO and Nd-LSCO the (underlying) 2D Fermi surface is hole-like, being centered at symmetry point Y = $(\frac{1}{2},\frac{1}{2})$ with $\dot{q} = 0.215 \pm 0.005$ r.l.u.\cite{42,43,44,45,46,47,48,49,50,51,52,53,54} The simultaneous occurrence of both metallic and insulating quantum states gives rise to a partial energy gap in the BZ, called ``pseudogap'' (see Fig. 2b).\cite{39,55,56}

With increased doping the region of metallic states on the Fermi surface (being a curve in 2D) widens about $\dot{Q}$, leaving a ``Fermi arc'' that bilaterally extends from $\dot{Q}$ out to the arc tips $\hat{Q}$ (see Fig. 2a).\cite{52,53,54} Quantum states along the remaining part of the underlying Fermi surface---from $\hat{Q}$ to the boundary of the BZ---are separated by the pseudogap $\tilde{\Delta} (q, x)$ depending on $q$-space position and doping.
It opens at $\hat{Q}$, where $\tilde{\Delta} (\hat{q}, x) \equiv 0$, and widens in an approximate square-root progression to a doping-dependent value $\tilde{\Delta} (0.5, x) \equiv \Delta^*(x)$ at the boundary of the BZ (see Fig. 2b).

As was noticed early on, at temperatures $T > 0$ the length of the Fermi arc qualitatively increases with doping $x$.\cite{52} For a quantitative assessment it is proposed here that the positions $\hat{Q}$ where the Fermi arc terminates---and the pseudogap opens---are determined by the incommensurability of the CDW, expressed, for $x > x_6$, by the condition for one of the lateral coordinates, say $q = q_a$,
\begin{equation}
\hat{q}(x) = \dot{q} + \delta_{c}(x) \, .  
\end{equation}
Thereby the CDW incommensurability, $\delta_c(x) = \hat{q}(x) - \dot{q}$, provides a rough measure for the length of (each wing of) the Fermi arc for a given doping level $x$ and at $T=0$. As $\delta_{c}(x)$ of Eq. (1) is derived in tetragonal approximation, $a_0 = b_0$, this approximation will be maintained. 

Experiments have shown that CDWs are \emph{unidirectional} in domains of the $CuO_2$ planes of LACO and Ln-LSCO (Ln = Nd, Eu) single crystals.\cite{34} For doping $x > x_6$, they are oriented along either the crystalline $a$ or $b$ direction in respective $a$-domains and $b$-domains. For $x < x_6$, they are oriented diagonally. (If the CDWs were oriented in two orthogonal directions within a domain, a checkerboard pattern would result---a possibility that was considered in early research but disproved by experiment.\cite{34}) As assumed in the derivation 

\pagebreak
\includegraphics[width=4.32in]{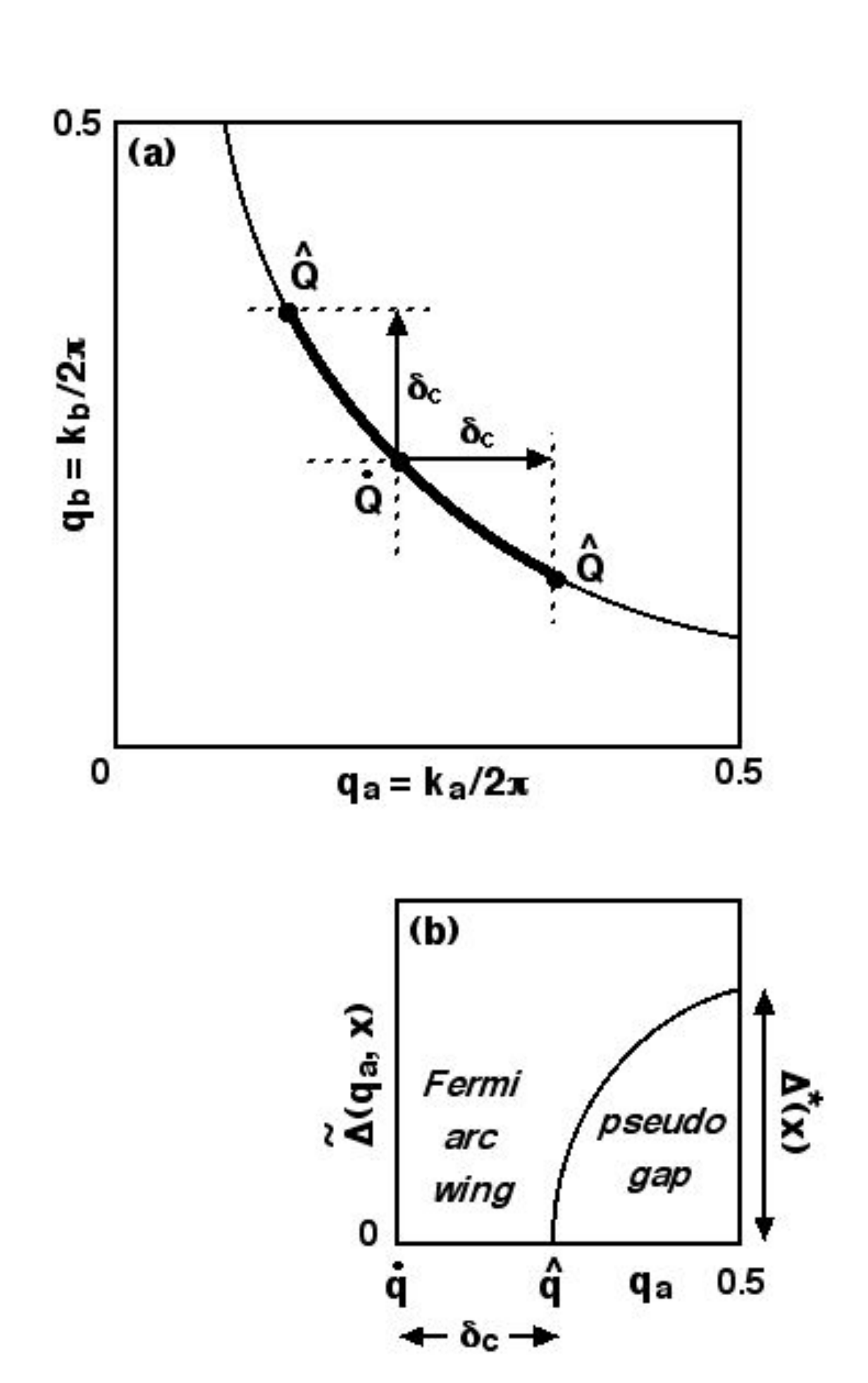}

\noindent FIG. 2. (a) First quadrant of the first Brillouin zone of $La_{2-x}Ae_{x}CuO_{4}$ ($Ae = Sr, Ba$) with (approximate) Fermi arc (bold) and pseudogap states along the underlying (unoccupied) Fermi surface (thin curve). Itinerant holes on each wing of the Fermi arc are trapped by unidirectional Bragg-reflection mirrors (dashed) whose extension equals the incommensurability of the CDW, $\delta_{c}(x)$, for a given doping level $x$. (b) Dependence (schematic) of the pseudogap $\tilde{\Delta}(q_a, x)$ on a lateral $q$-space coordinate from the center of the Fermi arc, $\dot{q}$, along a Fermi-arc wing to the wing tip, $\hat{q}$, and further along the underlying Fermi surface to the boundary of the Brillouin zone at $q_a=0.5$.

\pagebreak 

\noindent of Eq. (1), the CDW incommensurability can be considered the reciprocal value of the lattice constants $A_0(x) = B_0(x)$ of the doped-hole superlattice, $\delta_c(x)=A_0^{-1}(x)=B_0^{-1}(x)$.\cite{35} This makes $\delta_c(x)$ analogous to the reciprocal value of the crystal lattice constants, $q_{0a} = a_0^{-1}$ and $q_{0b} = b_0^{-1}$. The latter determine the size of the 2D BZ. Likewise $\delta_c(x)$ furnishes the extension of unidirectional CDW  Bragg-reflection mirrors, associated with corresponding $a$-domains and $b$-domains. Attached to $\dot{Q}$, the Bragg-reflection mirrors are spanned by vectors of length $\delta_c(x)$ parallel to the $q_a$ and $q_b$ axes for $x > x_6$ (as illustrated in Fig. 2a), but diagonally for $x < x_6$.  In this way the CDW Bragg-reflection windows trap itinerant holes on the Fermi-arc wings $\hat{Q}\dot{Q}$ and $\dot{Q}\hat{Q}$.

It needs to be pointed out that Eq. (2) holds only for the more important case of parallel stripes when $x > x_6$. The less important case when stripes are diagonal, $x < x_6$, is more complicated. If the same procedure of attaching $\delta_c(x)$ vectors to the Fermi dot $\dot{Q}$ is adopted, both a short and long Fermi-arc contribution results, caused by almost perpendicular and, respectively, parallel intersections of Bragg-reflection mirrors [orthogonal to the diagonal $\delta_c(x)$ vectors] with the underlying Fermi arc. It is not clear how to resolve this situation. Therefore the question of Fermi-arc length will left open for this doping range, $x_{10} < x < x_6$. Some insight can be gained, however, from the experimental data of the pseudogap temperature $T^*(x)$ in this doping range (see below).

It should be noted that not only the length of the Fermi arc, but also its \emph{curvature} changes with the doping level $x$ of LACO and Ln-LSCO as observed with ARPES and confirmed with Hubbard-model calculations.\cite{44,45,46,48,50}
Whereas the position of the Fermi dot $\dot{Q} = (\dot{q},\dot{q})$, is essentially constant, $\dot{q} = 0.215 \pm 0.005$, the curvature of the underlying Fermi surface, centered for low and moderate doping at symmetry point Y (hole-like), relaxes with increased doping until the Fermi arc reaches (with little curvature) the antinodal symmetry points M = $(\frac{1}{2},0)$ and $(0,\frac{1}{2})$, closing the pseudogap (see Fig. 4 below). 
This is the case at the doping level $x_0^* = 0.182 \pm 0.005$ for LACO but at $x_0^* = 0.235 \pm 0.005$ for Ln-LSCO.
With further increase of doping the Fermi arc keeps straightening and warping oppositely until it becomes centered at the origin of the BZ, $\Gamma = (0,0)$ (electron-like).

Due to less relaxation of curvature in Ln-LSCO the Fermi arc intercepts, at doping level $\overline{x}$, the boundary of the first BZ slightly away from (i. e., before) the M points and extends into the second BZ up to its termination by the Bragg-reflection mirror at $\hat{q} > 0.5$ (see Fig. 3a). Not only are the Fermi-arc segments in the \emph{second} BZ terminated at $\hat{Q}$ by Bragg-reflection mirrors, they are also, due to \emph{lattice} Bragg reflection, pinched off at $q = 0.5$ by the boundary of the lattice BZ---and thus isolated---from the main part of the Fermi arc in the first BZ. If the CDWs are strong enough in intensity, then it is conceivable that the pinched-off Fermi-arc segments in the corresponding parts of the other quadrants reconfigurate and join each other to form small rings---``Fermi ringlets'' would be a good expression---in the second BZ. The Fermi ringlets would be ``electron-like'' as they are centered at the origin of the BZ, $\Gamma$. Such Fermi ringlets may correspond to ``electron pockets'' in the Fermi surface accounting for for quantum oscillation in electron-doped lanthanide cuprates, $Ln_{2-x}Ce_xCuO_4$ ($Ln = Pr, Nd$) and hole-doped cuprates $YBa_2Cu_3O_{6+y}$, $YBa_2Cu_4O_8$, $HgBa_2CuO_{4+y}$
and based on negative Hall coefficients $R_H$ at low temperature.\cite{61,62,63,64,65,66,67,68,69,70,71}
Note that present model accounts only for the incommensurability, not the intensity of CDWs. Accordingly the presence of Fermi-arc tips in the second BZ, $\hat{q} > 0.5$, is a \emph{necessary}, not a sufficient condition for quantum oscillations.
However, if the CDWs are strong enough in intensity, then \emph{quantum oscillations} could be expected for the hole-doped Ln-LSCO (Ln = Nd, Eu) in the high-end doping interval of the pseudogap phase, $0.182 = \overline{x} < x < x_0^* = 0.235$.

\section{PSEUDOGAP}

An important situation is at hand when the tips of the Fermi arc, $\hat{Q}$, touch the boundary of the BZ, 
\begin{equation}
\hat{q}(\overline{x}) \equiv 0.5  \, .
\end{equation}
This is the case at a doping
level 
\begin{equation}
\overline{x} = 2(0.5 - \dot{q})^2 + x_{0}^N  \, ,  
\end{equation}
\noindent obtained by combining Eqs. (1) - (3).
Inserting $\dot{q} = 0.215 \pm 0.005$ and $x_0^N = 0.02$ gives $\overline{x} = 0.182 \pm 0.005$ for LACO and Ln-LSCO. A result from both increasing arc length and decreasing curvature with more doping, the Fermi arc eventually reaches the symmetry points M. This closes the pseudogap and contiguously joins the Fermi arcs of all quadrants of the first BZ to form a complete Fermi surface with attending metallicity. However, in the present case the metallicity is ``strange'' as explained below. The corresponding doping level, $x_0^*$, marks the high-doping end of the pseudogap phase at $T = 0$, often regarded a \emph{quantum critical point} (QCP).

\pagebreak
\includegraphics[width=5.82in]{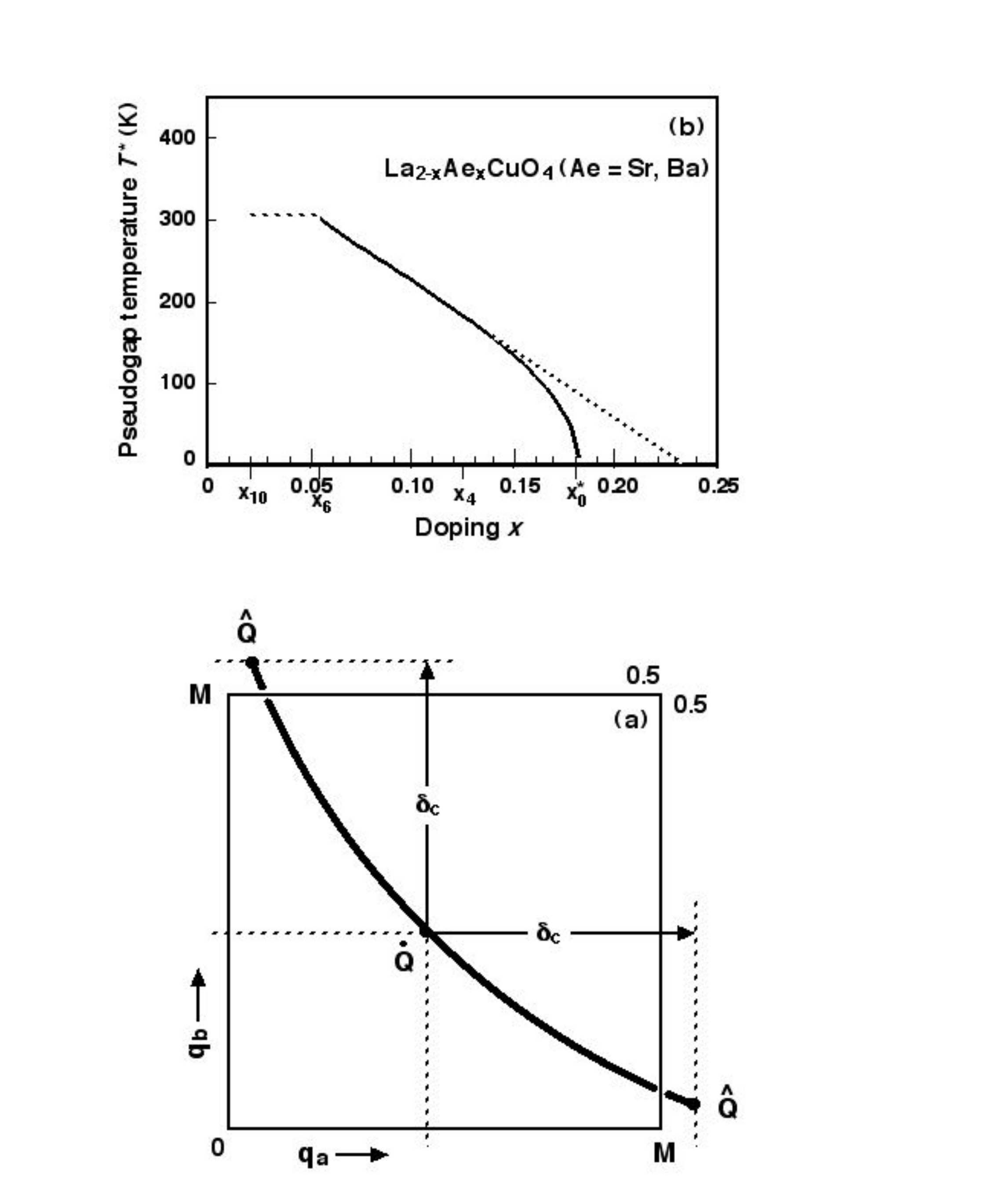}

\footnotesize

\noindent FIG. 3. (a, bottom) First quadrant of the first Brillouin zone with Fermi arc (schematic) extending beyond the BZ boundary. The pinched-off pieces in the second BZ may reconfigurate and combine with equivalent pieces in the other quadrants (not shown) to form ``Fermi ringlets'' in the second BZ, providing a necessary condition for quantum oscillations. 
(b, top) Doping dependence of the pseudogap temperature $T^*(x)$ of $La_{2-x}Ae_{x}CuO_{4}$ $(Ae = Sr, Ba)$ according to Eq. (8) (solid curve) in good agreement with experimental data. Quantum critical point $x_0^* = 0.182$ and linear extrapolation (slanted dotted line) to $x_{0\ell X}^*=0.235$. The horizontal hatched line indicates experimental values for $x < x_6$ (Ref. 72).

\normalsize

\pagebreak 

By connecting the $\delta_c(x)$ vectors to the Fermi dot $\dot{Q}$, the present model provides only the doping level $\overline{x}$ for the Fermi arc's touching the border of the BZ. It cannot furnish the QCP $x_0^*$ when the Fermi arc reaches the M points. Instead, $x_0^*$ must be obtained from experiment (extrapolated to T = 0) or from Hubbard-model calculations. In LACO, however, the Fermi arc reaches the boundary of the BZ \emph{at} the symmetry points M, as observed by ARPES and confirmed with Hubbard-model calculations.\cite{44,45,46,47,48} Thus in LACO the doping level for BZ boundary-touch and QCP \emph{coincide}, $\overline{x} = x_0^* =  0.182 \pm 0.005$. This is in contrast to Ln-LSCO where the Fermi arc first touches the boundary of the BZ---\emph{near} (not at!) points M with doping $\overline{x} = 0.182 \pm 0.005$---but, because of more Fermi-arc warping and after extension into the second BZ, reaches points M later at doping level $x_0^* = 0.235 \pm 0.005$.

For a derivation of the pseudogap temperature $T^*(x)$ that marks the boundary of the pseudogap phase and strange-metal phase we exploit, in the case of LACO, the coincidence $x_0^* = \overline{x}$. To this end we model the pseudogap energy function, at $T=0$, by a square-root dependence on one of the planar coordinates, $q$,
\begin{equation}
\tilde{\Delta} (q, x) = D \sqrt{q - \hat{q}(x)},
\end{equation}
\noindent where $D$ is a constant amplitude.
The value of the pseudogap at the BZ boundary is
\begin{equation}
\Delta^*(x) \equiv \tilde{\Delta} (0.5, x).
\end{equation}
At temperature $T>0$, closure of the pseudogap is provided by thermal energy,
\begin{equation}
k_BT^* = \gamma \Delta^*(x),
\end{equation} with $k_B$ being the Boltzmann constant and $\gamma = 0.465$ a scale factor.\cite{55,57,58} Combining Eqs. (4) to (7) gives the pseudogap temperature,
\begin{equation}
T^*(x) = C \sqrt{0.5 - \dot{q} - c^+\sqrt{x-x_{0}^N}}.
\end{equation}
\noindent Here the coefficient of the inner square root is $c^+ = w_c \, \Omega^+/4 = \sqrt{2}/2$ from Eq. (1). The coefficient of the outer square root is $C =\gamma D/k_B$. 
Figure 3b shows the doping dependence of the pseudogap temperature, Eq. (8), obtained with values of $\dot{q} = 0.215$ and $x_{0}^N = 0.02$, and $C$ fitted to experimental data of $T^*(x)$ in the high-temperature region (more about that shortly). 
The upper part of the curve in Fig. 3b is approximately linear in the range $0.056 \simeq x_6 < x < x_4 = 0.125$ but then turns down steeper to intercept the ordinate axis at $x_0^* = 0.182$. 
At a fixed doping level $x$, an increase of temperature in the pseudogap phase, $0<T<T^*$, while failing to close the pseudogap, correspondingly extends the length of Fermi arc, as has been observed.\cite{52,53,54}

Experimental $T^*(x)$ data depend, through the pseudogap amplitude $D$ in Eq. (5), on the probe and its sensitivity to detect the pseudogap phase.\cite{39,55,56,57,58,72,73,74}
In early measurements with ARPES, resistivity, Hall coefficient, susceptibility or heat capacity as probes, the pseudogap temperature extends from $T^*(0.05) \approx 700$ K diagonally across the phase diagram toward a linear extrapolation ($\ell X$) at $x_{0 \ell X}^* = 0.24 \pm 0.01$ and $T=0$.\cite{39,55,73} More recent measurements with such probes, but higher sensitivity, give pseudogap temperatures that taper across the phase diagram from $T^*(0.05) \approx 300$ K to the same linear extrapolation ($\ell X$) at $x_{0 \ell X}^* = 0.24 \pm 0.01$.\cite{57,58,72} The dashed $T^*(x)$ line in Fig. 1a is based on these newer data. Correspondingly the coefficient of proportionality, $C$, of the outer square root in Eq. (8) was fitted to the newer data in the scaling of the $T^*(x)$ curve in Fig. 3b.
If instead the Nernst signal (upturn of the Nernst coefficient) is used as a probe to determine $T^*(x)$, a linear dependence is observed again: 
In this case from from $T^*(0.05) \approx 200$ K to $T^*(0.20) \approx 50$ K and linear extrapolation to $x_{0 \ell X}^* = 0.26$.\cite{75,76}
Quite generally, for a given probe Eq. (8) yields the pseudogap temperature $T^*(x)$ of LACO by fitting the scaling coeffient $C$ to corresponding high-temperature data. 

An analysis of resistivity and Hall coefficient measurements inside the superconducting dome (with superconductivity destroyed by a sufficiently strong magnetic field) gives a steeper than linear descent for low temperature with values of $x^*_0 = 0.185 \pm 0.005$ for LSCO but $x^*_0 = 0.235 \pm 0.005$ for Ln-LSCO.\cite{60,77,78,79,80,81,82,83}  
With the steep downturn of the calculated $T^*(x)$ curve at low $T$ (see Fig. 3b) the value $x^*_0 = 0.182$ from Eq. (4) is in very good agreement with $x^*_0 = 0.185 \pm 0.005$ from resistivity and Hall coefficient experiments on LSCO.\cite{66,78,82,83}
Experimental data of the pseudogap temperature level off sharply in the doping range of diagonal stripes, $x_{10} < x < x_6$, to a value $T^*(x) \simeq T^*(x_6)$ as indicated by the horizontal hatched line in Fig. 3b.\cite{72} Interpreted in the spirit of Eq. (8) this would mean a constant length of Fermi arc in the very low doping range---a topic that needs more clarification.

Concerning $La_{2-x}Ba_{x}CuO_{4}$ there is a dearth of $T^*(x)$ data in the literature, presumably because of difficulties in crystal growth and experimentation.\cite{84} However, since the $La_{2-x}Ae_xCuO_4$ compounds
have the same doping dependence of the incommensurability of density waves $\delta_{c,m}(x)$, Eq. (1), and the same position of the Fermi dot $\dot{Q}$, it can be expected by Eq. (8) that the same $T^*(x)$ curve would hold for lanthanum cuprate doped with either alkaline-earth species, $Ae=Sr,Ba$.

\section{QCP DISCREPANCY FOR $\mathbf{La_{2-y-x}Ln_ySr_xCuO_4}$}

The value of $x^*_0 = 0.235$ for isostructural Ln-LSCO (Ln = Nd, Eu)---also of the same doping dependence of incommensurability $\delta_{c,m}(x)$ and the same position of Fermi dot $\dot{Q}$---is surprising at first glance.\cite{75}
Looking for possible reasons, the most likely cause should originate with the most basic difference: The different electron configuration of the substituting $Nd^{3+}$ ions ($4f^35s^2p^6$) or $Eu^{3+}$ ions ($4f^65s^2p^6$) compared to the substituted $La^{3+}$ ions ($4f^05s^2p^6$) with magnetic moments of $\mu(Nd^{3+})=4.04$ $\mu_B$  and $\mu(La^{3+})=0$
[or $\mu(Nd^{3+}) = 3.5$ $\mu_B$ and $\mu(Eu^{3+}) = 3.4$ $\mu_B$ in paramagnetic circumstances],\cite{2,85} giving rise to additional magnetic order in Ln-LSCO.
Circumstantial support for magnetic effects from $Nd$ or $Eu$ can be inferred from magnetic order in elemental $Nd$ and $Eu$ metal,\cite{85} in $La_{1.65}Nd_{0.35}CuO_4$ and $La_{1.48}Nd_{0.40}Sr_{0.12}CuO_4$ compounds,\cite{2,86,87} and possibly from the much lower superconducting transition temperature $T_c(x)$ in Nd-LSCO than in LSCO.\cite{3} Whatever the extra magnetic order from $Nd$ or $Eu$ substitution, it does \emph{not} affect the incommensurability $\delta_c(x)$---and thus the \emph{size} of the CDW Bragg-reflection mirrors---nor the collapse of 3D-AFM, as inferred from the dependence of $\delta_c(x)$ on $x_0^N$, Eq. (1). On the other hand, the extra magnetic order from $Ln$ substitution seems to affect the doping dependence of the \emph{curving} of the Fermi arc. Energy distribution curves (EDCs) of $La_{1.6-x}Nd_{0.4}Sr_xCuO_4$, obtained with ARPES, are similar to those of $La_{2-x+\Delta x}Sr_{x-\Delta x}CuO_4$, for $x = 0.15, 0.20, 0.24$ with $\Delta x \approx 0.05$.\cite{51} 
In other words, the EDC evolution of Nd-LSCO in that doping range lags behind that of $x$-doped LSCO by $\Delta x \approx 0.05$.

\section{STRANGE METAL}
When the Fermi arc reaches the M points in $q$-space, the pseudogap closes, causing the formation of a contiguous Fermi surface across the entire BZ. The compound then is \emph{fully} metallic with an observed jump in itinerant hole density to $1+p$ from previously $p=x$.\cite{81,82} The additionally liberated holes---one hole per unit cell---don't participate in the CDW, only the \emph{doped} (excess) holes do. Although fully metallic, this phase is \emph{not} a Fermi liquid, hence 

\pagebreak
\includegraphics[width=8in]{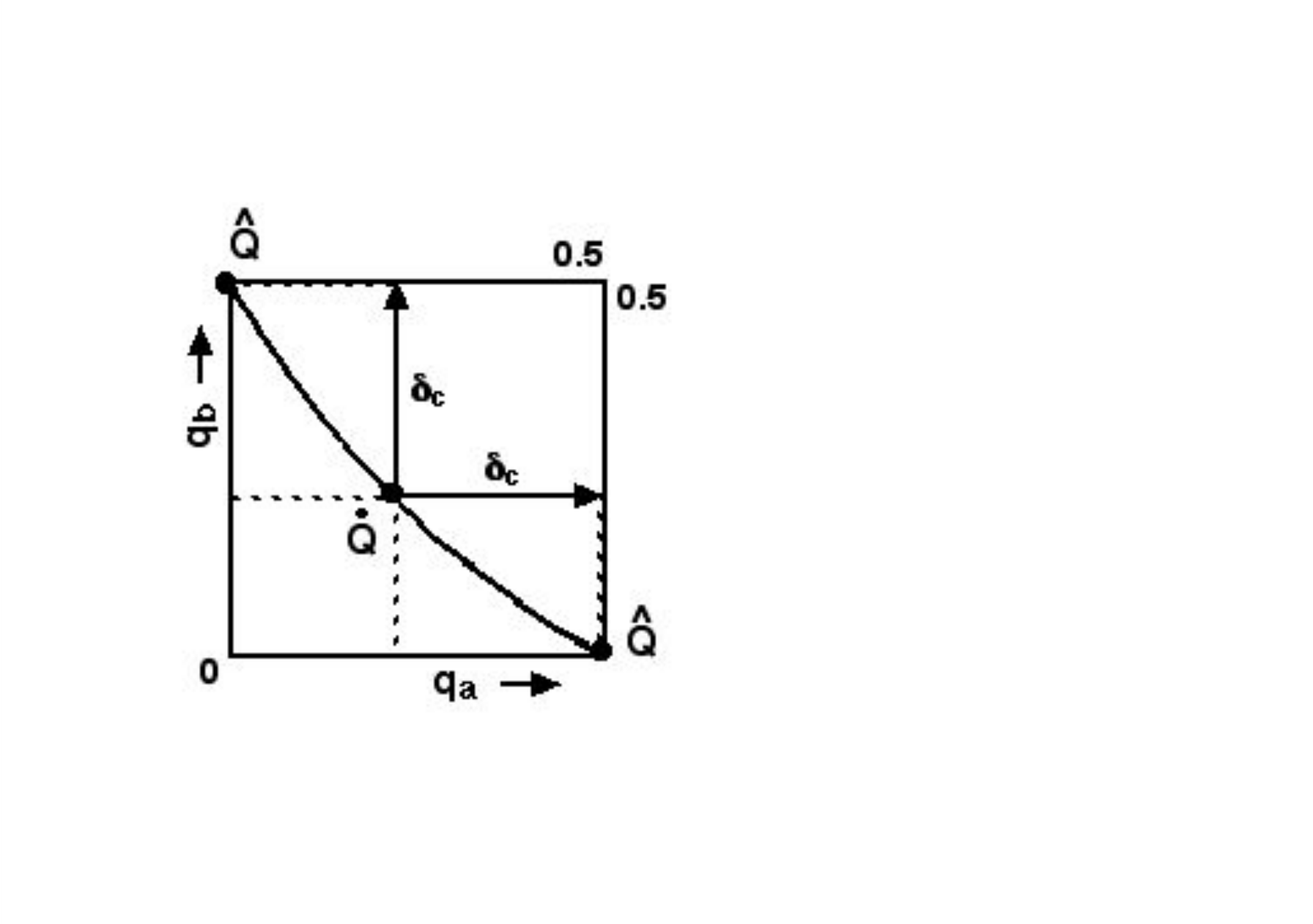}

\noindent FIG. 4. At doping level $x_0^*$ and temperature $T=0$ (quantum critical point) the Fermi arc (schematic and for LACO) extends contiguously across the entire first Brillouin zone (shown here: first quadrant only) causing a closing of the pseudogap. Conflicting Bragg-reflection conditions at the boundaries of the BZ (solid line) and the CDW Bragg-reflection mirrors (dashed line) frustrate \textit{umklapp} processes of carrier-carrier scattering which manifests the strange-metal phase.

\pagebreak

\noindent called \textit{strange-metal} phase.

The resistivity of a Fermi-liquid metal is proportional to the square of the temperature, $\rho \propto T^2$, as a result of charge carrier-carrier scattering which necessitates \emph{umklapp} processes.\cite{88} 
An \emph{umklapp} process is the back-reflection of the vector sum of two charge carriers' momentum in a collision (here, divided by $h$), $\mathbf{q_3} = \mathbf{q_1} + \mathbf{q_2}$, from the second BZ into the first by subtraction of a reciprocal lattice constant, $\mathbf{q'_3} = \mathbf{q_3} - \mathbf{q_0}$. The observed ``strange-metal'' phase that commences beyond the pseudogap phase is characterized by a \emph{linear} temperature dependence of the resistivity, $\rho \propto T$, resulting from mere charge carrier-phonon scattering.\cite{88} 
Why is carrier-carrier scattering missing? When the CDW Bragg-reflection mirrors extend to or beyond the boundary of the BZ, conflicting Bragg conditions hold: both a reflection by $\Delta q = q_0 = \pm 1$ (r.l.u.) to the opposite boundary of the BZ \emph{and} a reflection by $\Delta{q} = \delta_c(x_0^*)$ [$= \pm 0.5$ (r.l.u.) for LACO] to the opposite end of the CDW reflection mirrors (see Fig. 4).  The conflict frustrates the \emph{umklapp} process, giving rise to strange-metal behavior.

\bigskip \bigskip 

\centerline{ \textbf{ACKNOWLEDGMENTS}}

\noindent I thank Duane Siemens for valuable discussions and Preston Jones for help with LaTeX. 
\linebreak I also thank M. R. Norman for information on the probe dependence of $T^*$.

\end{document}